\documentstyle[aps,pra,epsfig,twocolumn]{revtex}

\def\be{\begin{equation}}
\def\ee{\end{equation}}
\def\bea{\begin{eqnarray}}
\def\eea{\end{eqnarray}}
\def\bma{\begin{mathletters}}
\def\ema{\end{mathletters}}

\def\C{\hbox{$\mit I$\kern-.7em$\mit C$}}

\begin{document}
\draft

\title{Multipartite bound entangled states that violate Bell's inequality}

\author{W. D\"ur}

\address{Institut f\"ur Theoretische Physik, Universit\"at Innsbruck,
A-6020 Innsbruck, Austria}

\date{\today}

\maketitle

\begin{abstract}
We study the relation between distillability of multipartite
states and violation of Bell's inequality. We prove that there
exist multipartite bound entangled states (i.e. non-separable,
non-distillable states) that violate a multipartite Bell
inequality. This implies that (i) violation of Bell's inequality
is not a sufficient condition for distillability and (ii) some
bound entangled states cannot be described by a local hidden
variable model.
\end{abstract}


\narrowtext

%-------------------------------------------------------------------------
%-------------------------------------------------------------------------
 %Introduction
%-------------------------------------------------------------------------
%-------------------------------------------------------------------------
 % define: distillable, violation of Bell inequality, separable, entangled, bound entangled, PPT/NPT !!!

Since the celebrated work by Bell \cite{Be64}, it is evident that
quantum mechanics is not compatible with local realist theories.
In fact, Gisin \cite{Gi91} and Gisin and Peres \cite{Gi92} showed
that all bipartite entangled pure states violate the Bell-CHSH
inequality \cite{Cl69}. This rules out the existence of a local
hidden variable (LHV) model which is capable of describing all
statistical correlations predicted by quantum mechanics for such
states. These results were readily generalized by Popescu and
Rohrlich \cite{Po92} to multipartite entangled pure states of
arbitrary dimension.

However, for quantum systems in mixed states, the situation is
much more complicated and we still lack a complete classification
of mixed states into ``local'' and ``non--local'' ones. Although
structural knowledge about entanglement has increased rapidly in
the last decade, many open questions remain to be answered, in
particular concerning the relation of certain entanglement
properties to the existence of LHV models.

While it is obvious that separable mixed states \cite{notSep} can
be described by a LHV model, Werner showed in his pioneering paper
\cite{We89} that also a certain class of {\it entangled} mixed
states ---now called Werner states--- do not violate any
Bell--type inequality. This was done by explicitly constructing a
LHV model which can simulate the results of any single (i.e.
nonsequential) measurement performed at each side. However, some
years later Popescu \cite{Po95} realized that also sequential
measurements can be considered, and showed that most of the Werner
states exhibit violation of local realism if sequences of
measurements are taken into account. This so called ``hidden
nonlocality'' is revealed by a sequence of two measurements, where
the first measurement is used to select a certain subensemble of
pairs ---those pairs which produce a specific outcome--- while the
second measurement tests the Bell observable on the subensemble.
If the subensemble does not satisfy Bell's inequality, then one
concludes that the initial ensemble violates local realism. In
fact, applying a similar reasoning to collective tests of
particles \cite{Pe96b}, it was shown that all inseparable states
of two qubits \cite{Ho97} violate local realism. More generally,
all {\it distillable} \cite{noteDist,Be96} states violate local
realism. It is an open problem whether violation of Bell's
inequality ---which seems to be a rather strong requirement---
already implies distillability \cite{notePe99}.

While in systems of two qubits all entangled states are
distillable \cite{Ho97} (and thus violate local realism), this
turned out to be false for higher dimensional systems. The
Horodecki \cite{Ho98} discovered states which, although being
non--separable and thus entangled, are not distillable. Those
states are called bound entangled. The role of bound entangled
states is not entirely clear yet. Although they can not be useful
for any quantum information processing task directly (since they
are non--distillable), it was nevertheless shown that they allow
to perform certain processes (e.g. quasidistillation) \cite{Ho99}
which cannot be performed using local operations and classical
communication alone. In particular, it is not known whether bound
entangled states violate local realism or not. This paper is aimed
to shine some light on these questions.

We will, however, tackle this problem not in its bipartite setting
but rather in the multipartite setting. To this aim, we consider
$N$ spatially separated parties. A generalization of the Bell-CHSH
inequality to multipartite systems is due to Mermin \cite{Me90},
and was further developed e.g. in Ref. \cite{Bel93} (see also
Refs. \cite{Gi98a,We01,Zu01,Gi01} for some recent developments).
Also the notion of separability and distillability can be readily
generalize to multipartite systems (see e.g. Refs.
\cite{Du00A,Du00BE}). We have that a $N$--partite state $\rho$ is
called (fully) separable iff it can be written as a convex
combination of (unnormalized) product states, i.e. \be \rho=\sum_i
|a_i\rangle_{\rm party1}\langle a_i| \otimes |b_i\rangle_{\rm
party2}\langle b_i|\otimes \ldots \otimes |n_i\rangle_{{\rm party}
N}\langle n_i|.\label{sepa1} \ee
If $\rho$ is not fully separable,
it is entangled. Note that fully separable states are those which
can be prepared locally. Due to the fact that many different kinds
of multipartite pure state entanglement exist, there are various
kinds of distillability \cite{Du00BE}. We will however not
distinguish between those possible kinds of entanglement but
rather say that a state $\rho$ is distillable, iff ---by means of
local operations and classical communication--- out of an
arbitrary number of identical copies of a state $\rho$ {\it some}
entangled pure state can be created. If no entangled pure state
whatsoever can be created, $\rho$ is non--distillable. As shown in
Ref. \cite{Du00A} (see also \cite{Sm00}), there exist also
multipartite bound entangled states, i.e. states which are not
(fully) separable and hence entangled, but which are
non--distillable. Here we investigate multipartite bound entangled
states and consider the relation of their distillability to the
existence of a LHV model. We show that there exist multipartite
bound entangled states which violate Bell's inequality. This means
on the one hand that violation of Bell's inequality does not imply
distillability, and on the other hand that some bound entangled
states violate local realism.

In the remainder of this letter, we consider specific $N$--qubit
mixed states $\rho_N$ acting on a Hilbert space ${\cal
H}=(\C^2)^{\otimes N}$. We consider $N$ parties, $A_1,\ldots,A_N$,
at different locations, each of them possessing several qubits.
The parties possess $M$ identical copies of $\rho_N$, where $M$
can be arbitrary large. Thus the state of all qubits is described
by the density operator $\rho_N^{\otimes M}$. This ensures that
the parties can use distillation protocols \cite{Be96} in order to
obtain entangled pure states between some of them. The states
$\rho_N$ we consider are given by \cite{noterhoN}
\bea
\rho_N=\frac{1}{N+1}\left(|\Psi\rangle\langle
\Psi|+\frac{1}{2}\sum_{k=1}^{N}(P_k+\bar
P_{k})\right).\label{rhoN}
\eea
We have that $|\Psi\rangle$ is a $N$-party
Greenberger-Horne-Zeilinger (GHZ) state \cite{GHZ},
\be
|\Psi\rangle=\frac{1}{\sqrt{2}}(|0^{\otimes
N}\rangle+e^{i\alpha_N}|1^{\otimes N}\rangle) \label{GHZ},
\ee
with $\alpha_N$ being an arbitrary phase. We denoted by $P_k$ a
projector on the state $|\phi_k\rangle$, where  $|\phi_k\rangle$
is a product state which is $|1\rangle$ for party $A_k$ and
$|0\rangle$ for all other parties, i.e.
$|\phi_k\rangle=|0\rangle_{A_1}|0\rangle_{A_2}\ldots|1\rangle_{A_k}\ldots|0\rangle_{A_{N-1}}|0\rangle_{A_N}.$
The projector $\bar P_k$ is obtained from $P_k$ by replacing all
zeros by ones and vice versa.

For the states $\rho_N$ (\ref{rhoN}) we show

\begin{description}

\item [(i)] The states $\rho_N$ are bound entangled, i.e. non-separable and
non-distillable if the number of parties $N\geq 4$.

\item[(ii)] The states $\rho_N$ violate the Mermin-Klyshko inequality if the number of
parties $N\geq8$ and can thus not be described by a LHV model.

\end{description}

We start out by showing ${\bf (i)}$, i.e. $\rho_N$ is bound
entangled. One readily verifies that $\rho_N^{T_{A_k}}$ is a
positive operator $\forall k$, where $T_{A_k}$ denotes partial
transposition with respect to party $A_k$ \cite{notePT}. This
already implies that $\rho_N$ is non--distillable \cite{Du00BE}.
To see this, assume on the opposite that one can distill some bi-
or multipartite entangled pure state. As shown below (Lemma 1),
one can always create by means of local operations from any
multipartite entangled pure state a maximally entangled {\it
bipartite} pure state shared between two of the parties, say $A_i$
and $A_j$. Thus the resulting state has non-positive partial
transposition with respect to parties $A_i$ and $A_j$, while the
initial state $\rho_N$ has positive partial transposition. Due to
the fact that by means of local operations and classical
communication, one cannot change the positivity of the partial
transposition \cite{Ho99}, we have the desired contradiction,
hence $\rho_N$ is not distillable.

On the other hand, $\rho_N$ can easily seen to be entangled for
$N\geq 4$, e.g. by observing that $\rho_N^{T_{A_kA_l}}$ is not a
positive operator for $k\not=l$ and $N\geq 4$. This already
implies that $\rho_N$ is not fully separable, as positivity of all
possible partial transpositions is a necessary condition for
(full) separability in multipartite systems \cite{Pe96,Du00A}. It
remains to show the announced

%-------------------------------------------------------------------------
%-------------------------------------------------------------------------
 %proof that one can create bipartite entanglement from multipartite entanglement:
%-------------------------------------------------------------------------
%-------------------------------------------------------------------------

{\it Lemma 1:} Given a single copy of an $N$--party entangled pure
state $|\Phi\rangle$ of arbitrary dimension, one can always create
with non--zero probability of success by means of local operations
a maximally entangled bipartite pure state shared among some of
the parties.
\\
{\it Proof:} One may write $|\Phi\rangle$ in its Schmidt
decomposition with respect to any party $A_k$. Since
$|\Phi\rangle$ is entangled, there exist at least one party, say
$A_1$, where one obtains a minimal number of two non-zero Schmidt
coefficients (otherwise $|\Phi\rangle$ would be a $N$-party
product state). That is,
\be
|\Phi\rangle=\sum_{k=0}^d\lambda_k
|k\rangle_{A_1}|\varphi_k\rangle_{A_2\ldots A_N}, \ee where
$\langle k| k' \rangle =\langle \varphi_k| \varphi_{k'} \rangle
=\delta_{kk'}$ and $d\geq2$. For simplicity, let us assume that
$d=2$ and $\lambda_0=\lambda_1=1/\sqrt{2}$ (this can always be
accomplished by a filtering measurement in $A_1$, e.g. using
$O_{A_1}=\lambda_1|0\rangle\langle 0|+\lambda_0|1\rangle\langle
1|$). Now, either (a) both $|\varphi_0\rangle$ and
$|\varphi_1\rangle$ are product states, or (b) at least one of
them, say $|\varphi_0\rangle$, is entangled.

In case of (a), $|\varphi_0\rangle$ and $|\varphi_1\rangle$ have
to be locally orthogonal in at least one location, say at $A_2$
(this is due to the fact that $\langle \varphi_0|
\varphi_{1}\rangle=0$), i.e.
$|\varphi_0\rangle=|0\rangle_{A_2}|\chi_3\rangle_{A_3}\ldots
|\chi_N\rangle_{A_N}$ and
$|\varphi_1\rangle=|1\rangle_{A_2}|\tilde\chi_3\rangle_{A_3}\ldots
|\tilde\chi_N\rangle_{A_N}$. There may exist $l$ locations $A_k$,
$l\leq N-2$ for which $|\chi_k\rangle=|\tilde\chi_k\rangle$. Each
of the other parties $A_k$ can apply a local filtering measurement
of the form
$O_{A_k}=|0\rangle\langle\chi_k'|+|1\rangle\langle\tilde\chi_k'|$,
where $\{|\chi_k'\rangle,|\tilde\chi_k'\rangle\}$ is the
biorthonormal basis to $\{|\chi_k\rangle,|\tilde\chi_k\rangle\}$,
i.e.
$\langle\tilde\chi_k|\chi_k'\rangle=\langle\chi_k|\tilde\chi_k'\rangle=0$
and
$\langle\tilde\chi_k|\tilde\chi_k'\rangle=\langle\chi_k|\chi_k'\rangle=1$.
One readily observes that this leads to the creation of a $N-l$
party GHZ state (\ref{GHZ}), from which ---by means of local
measurements in the basis $\{|+\rangle,|-\rangle\}$, where
$|\pm\rangle=1/\sqrt{2}(|0\rangle\pm|1\rangle)$ at the remaining
locations--- a maximally entangled bipartite pure state shared
between any two out of the remaining $N-l$ parties can be created
deterministically.

In case of (b), one measures in $A_1$ the projector
$P_{A_1}=|0\rangle\langle 0|$ and is left with an entangled state
of $N-1$ (or less) particles. This situation is similar to the one
we started with, however the number of entangled systems
decreased. One proceeds in the same vain with the remaining
systems until (a) applies, which happens in the worst case if only
two entangled particles are left. Finally, one obtains at least a
maximally entangled bipartite state shared among two of the
parties which concludes the proof of the lemma.

Note that from Lemma 1 follows the non--existence of a LHV model
for {\it all} multipartite entangled pure states of arbitrary
dimension which describes properly also sequences of measurements
(see Ref. \cite{Po92} for the stronger result of inconsistency
with local realism even for single measurements per site).
Following the reasoning of Popescu \cite{Po95} (see also
\cite{Zu98}) (adopted to the multipartite case), the violation of
Bell's inequality of a subensembles of states (obtained e.g. by
local filtering measurements) ensures that also the original
ensemble violates local realism. Since, according to Lemma 1, from
any multipartite entangled pure state a maximally entangled
bipartite pure state can be created ---which clearly maximally
violates Bell's inequality--- the claim follows.

%-------------------------------------------------------------------------
%-------------------------------------------------------------------------
 %rho_N violates BI
%-------------------------------------------------------------------------
%-------------------------------------------------------------------------

We now turn to {\bf (ii)} and show that $\rho_N$ violates the
Mermin-Klyshko inequality for $N\geq8$.  Let $a_j,a'_j$ be two
vectors on the unit sphere which indicate two possible measurement
directions for party $A_j$. The corresponding observables are
given by $O_j=\sigma_{a_j},O'_j=\sigma_{a'_j}$. Up to a
normalization factor, any $k$--qubit Bell inequality involving two
observables per qubit can be written as
\be
|\langle {\cal B}_k \rangle | \leq 1 \label{BI}, \ee where ${\cal
B}_k\equiv{\cal B}_k(a_1,a_2,\ldots,a_k,a'_1,a'_2,\ldots,a'_k)$ is
the corresponding Bell operator. We consider the Mermin-Kyshko
inequalities \cite{Me90,Bel93} for $N$ qubits, whose corresponding
Bell operator is defined recursively as \cite{Gi01}
\be
{\cal B}_k=\frac{1}{2} {\cal B}_{k-1} \otimes (\sigma_{a_k}+\sigma_{a'_k}) + \frac{1}{2} {\cal B'}_{k-1} \otimes (\sigma_{a_k}-\sigma_{a'_k}),
\ee
where ${\cal B'}_k$ is obtained from ${\cal B}_k$ by exchanging all the $a_k$
and $a'_k$.

We choose the same measurement directions in all $N$ locations,
$\sigma_{a_j}=\sigma_x$ and $\sigma_{a'_j}=\sigma_y \forall j$, where
$\sigma_x,\sigma_y$ are Pauli matrices. One readily verifies that in this case
${\cal B}_N$ can be written as
\be
{\cal B}_N=2^{\frac{N-1}{2}}(e^{i\beta_N}|1^{\otimes N}\rangle\langle 0^{\otimes N}|+e^{-i\beta_N}|0^{\otimes N}\rangle\langle 1^{\otimes N}|),
\ee
with $\beta_N\equiv\pi/4(N-1)$. Using that tr$({\cal B}_N P_k)$ = tr$({\cal B}_N \bar
P_k) = 0 \forall k$ and tr$({\cal B}_N
|\Psi\rangle\langle\Psi|)=2^{\frac{N-1}{2}}$ when fixing the phase
$\alpha_N=\beta_N$ in $|\Psi\rangle$ (\ref{GHZ}), we have
\be
{\rm tr}({\cal B}_N \rho_N)=\frac{1}{N+1}2^{\frac{N-1}{2}},
\ee
which fulfills ${\rm tr}({\cal B}_N \rho_N) > 1$ iff $N\geq 8$. Thus the states
$\rho_N$ (\ref{rhoN}) with the choice $\alpha_N=\pi/4(N-1)$ violate the
Mermin-Klyshko inequality for $N\geq8$ as announced.

To conclude, we have shown that certain multipartite bound entangled states 
violate a multipartite Bell inequality. This implies that (i) violation of 
Bell's inequality is not a sufficient condition for distillability and (ii) 
there does not exist a local hidden variable model for certain bound entangled 
states. Note that the states $\rho_N$ for sufficiently large $N$ violate the 
Mermin-Klyshko inequality directly, and no sequence of measurements, eventually 
performed on a tensor product of the states, is required to rule out the 
existence of a LHV model as in the case of ``hidden nonlocality''. There remain 
a number of open problems concerning the relation of inseparability to the existance 
of a LHV model \cite{Te00}. In particular, it is not known whether all bipartite 
bound entangled states, those with positive partial transposition as well as the 
conjectured ones with non--positive partial transposition \cite{Di00}, can be 
described by a local hidden variable model or not.

%-------------------------------------------------------------------------

I thank G. Vidal for an essential contribution in the proof of
Lemma 1, and J. I. Cirac and G. Giedke for interesting
discussions. This work was supported by the Austrian SF under the
SFB ``control and measurement of coherent quantum systems''
(Project 11), the European Community under the TMR network
ERB--FMRX--CT96--0087 and project EQUIP (contract IST-1999-11053),
the ESF, and the Institute for Quantum Information GmbH.

%-------------------------------------------------------------------------
%-------------------------------------------------------------------------

%-------------------------------------------------------------------------
%-------------------------------------------------------------------------

%-------------------------------------------------------------------------
%-------------------------------------------------------------------------


\begin{references}
\bibitem{notSep}
A bipartite mixed state of two parties $A$ and $B$ is said to be separable iff
it can be written as a convex combination of product states, $\rho=\sum p_i
|a_i\rangle_A\langle a_i|\otimes |b_i\rangle_B\langle b_i|$. Otherwise, $\rho$
is entangled.
\bibitem{noteDist}
A bipartite state $\rho$ is called distillable, iff ---by means of local
operations and classical communication--- out of an arbitrary number of
identical copies of the state a maximally entangled bipartite pure state, e.g.
$|\Phi^+\rangle=1/\sqrt{2}(|00\rangle+|11\rangle)$, can be created.
\bibitem{notePe99}
This seems to be the intuition of Peres, who conjectures in Ref.
\cite{Pe99} that all states with positive partial transposition
can be described by a LHV model. Results in favour of this
conjecture were recently obtained in Ref. \cite{We99}.
\bibitem{noterhoN}
Note that $\rho_N$ belongs ---up to an irrelevant phase factor--- to a family of
$N$--partite mixed states studied in detail in Refs. \cite{Du00A,Du00BE}. The
corresponding parameters are given by $\lambda_0^+=1/(N+1), \lambda_0^-=0$
and $\lambda_k=1/(2N+2)$ if $k$ corresponds to a bipartite splitting one party - rest
and $\lambda_k=0$ otherwise.
\end{references}
\end{document}